\documentclass[article]{jss}

\usepackage{amsmath}

\usepackage[english]{babel}
\usepackage[parfill]{parskip}
\usepackage{afterpage}
\usepackage{framed}
\usepackage{xspace}

\usepackage{graphicx}
\usepackage[labelfont=bf]{caption}
\usepackage[format=hang]{subcaption}

\usepackage{booktabs}

\DeclareRobustCommand{\parhead}[1]{\textbf{#1}~}

\usepackage{fancyvrb}
\fvset{fontsize=\normalsize}

\usepackage{natbib}

\usepackage[acronym,smallcaps,nowarn]{glossaries}

\newacronym{SGD}{sgd}{stochastic gradient descent}
\newacronym{AISGD}{ai-sgd}{averaged implicit stochastic gradient descent}
\newacronym{ESGD}{\textup{explicit} sgd}{}
\newacronym{ISGD}{\textup{implicit} sgd}{}
\newacronym{GLM}{GLM}{generalized linear model}
\newacronym{M}{\textup{M-estimator}}{}
\newacronym{ASGD}{\textup{averaged} sgd}{}
\newacronym{CM}{cm}{classical momentum}
\newacronym{NAG}{nag}{Nesterov's accelerated gradient}

\glsunset{ESGD} 
\glsunset{ISGD}
\glsunset{ASGD}
\glsunset{M}

\renewcommand{\b}[1]{\mathbf{#1}}
\newcommand{\m}[1]{\b{#1}}
\newcommand{\bigO}[1]{\mathcal{O}(#1)}
\newcommand{\Ex}[1]{\mathrm{E}(#1)}

\newcommand{\scoren}[2]{\nabla \log f(\b{y}_{#2}; \b{x}_{#2}, #1)}

\newcommand{\thetastar}{\theta_\star}
\newcommand{\vtheta}{\b{\theta}}

\newcommand{\vthetan}{\b{\theta}_n}

\newcommand{\vthetanprev}{\b{\theta}_{n-1}}

\newcommand{\thetaim}[1]{\theta_{#1}^{\mathrm{im}}}
\newcommand{\thetaBar}[1]{\bar{\theta}_{#1}}

\newcommand{\vthetastar}{\b{\thetastar}}

\newcommand{\thetaimpl}[1]{\vtheta_{#1}^{\mathrm{im}}}
\newcommand{\thetasgd}[1]{\vtheta_{#1}^{\mathrm{sgd}}}

\newcommand{\x}{\b{x}}

\newcommand{\xn}{\b{x}_n}
\newcommand{\D}{\mathbf{D}}
\newcommand{\Y}{\mathbf{Y}}
\newcommand{\X}{\mathbf{X}}
\newcommand{\G}{\mathbf{G}}
\newcommand{\gnk}{\mathbf{g}_{nk}}

\newcommand{\yn}{\b{y}_n}

\newcommand{\Fisher}[1]{\boldsymbol{\mathcal{I}}(#1)}

\newcommand{\Reals}[1]{\mathbb{R}^{#1}}

\renewcommand{\Ex}[1]{\mathbb{E}\left (#1 \right )} 		 
\newcommand{\ExCond}[2]{\mathbb{E}\left (#1 \right|#2)} 		 
\newcommand{\Var}[1]{\mathrm{Var}\left(#1\right)}   		 

\usepackage{amssymb, amsthm, mathtools}
\usepackage{algorithm,algpseudocode}
\newtheorem{assumption}{Assumption}[section]

\newtheorem{theorem}{Theorem}[section]

\usepackage{color}

\author{Dustin Tran \\Harvard University \And
        Panos Toulis \\Harvard University \And
        Edoardo M. Airoldi \\Harvard University}
\title{Stochastic gradient descent methods \\ for estimation with large data sets}

\Plainauthor{Dustin Tran, Panos Toulis, Edoardo M. Airoldi}
\Plaintitle{Stochastic gradient decent methods for estimation with large data sets}
\Shorttitle{Package \pkg{sgd} for estimation with large data sets}

\Abstract{
We develop methods for parameter estimation in settings with large-scale data sets, where traditional methods are no longer tenable.
Our methods rely on stochastic approximations, which are computationally
efficient  as they maintain one iterate as a parameter estimate,
and successively update that iterate based on a single data point.
When the update is based on a noisy gradient, the stochastic approximation
is known as standard {\em stochastic gradient descent}, which
has been fundamental in modern applications with large data sets.
Additionally, our methods are numerically stable because
they employ {\em implicit} updates of the iterates. Intuitively,
an implicit update is a shrinked version of a standard one, where
the shrinkage factor depends on the observed Fisher information at the corresponding data point.
This shrinkage prevents numerical divergence of the iterates, which can be caused either by excess noise or outliers.
Our \pkg{sgd} package in \proglang{R} offers the most extensive and robust implementation of
stochastic gradient descent methods. We
demonstrate that \pkg{sgd} dominates alternative software in runtime for several estimation problems with massive data sets.
Our applications include the wide class of
generalized linear models as well as M-estimation for robust regression.
}
\Plainkeywords{stochastic gradient descent, massive data, exponential
family, GLMs, M-estimation}

\Address{
  Dustin Tran, Panos Toulis, Edoardo M. Airoldi\\
  Department of Statistics\\
  Harvard University\\
  1 Oxford Street, Cambridge, MA 02138, USA\\
  E-mail: \email{dtran@g.harvard.edu}, \email{ptoulis@fas.harvard.edu},
  \email{airoldi@fas.harvard.edu}
}

\begin{document}

\section{Introduction}
\label{sec:introduction}
Massive data sets as well as streaming data, in which one observes only a group
of data points at a time, are becoming increasingly common in modern
statistical analysis. Under the setting of hundreds of millions of observations
and hundreds or thousands of covariates \citep{Council:2013kx}, it becomes difficult to
estimate the parameters of a statistical model;
the three ideal properties are computational efficiency, statistical
optimality, and numerical stability, and it is challenging to address all three
with a single estimation method.

More formally, suppose there exists a vector of parameters $\vthetastar \in
\Reals{p}$ and that we observe i.i.d. samples $\D = \{\xn, \yn\}$, for
$n=1,2,\ldots,N$; in the $n^{th}$ data point $(\xn, \yn)$,
the outcome $\yn \in \Reals{d}$ is distributed
conditional on covariates $\xn \in \Reals{p}$ according to a
known density $f(\yn; \xn, \vthetastar)$, and thus the log-likelihood function for the entire data set $\D$ is given by $\ell(\theta; \D) = \sum_{n=1}^N \log f(\yn; \xn, \theta)$.
The task is to estimate the true parameter value $\thetastar$ when $N$ is
infinite (streaming setting), or to approximate some estimator of $\thetastar$, such as the maximum-likelihood estimator
$\theta^{\mathrm{mle}} = \arg \max_{\theta \in \Reals{p}}\ell(\theta; \D)$,
when $N$ is finite.

Widely used methods for statistical estimation, such as Fisher scoring, the EM algorithm, and iteratively reweighted
least squares \citep{Fisher:1925vn,Demp:Lair:Rubi:1977,green1984iteratively}
are not feasible in such settings; either they strictly do not apply
in the streaming setting (infinite $N$), or they do not scale to large data (finite but large $N$).
Fisher scoring, for example, requires at each iteration the inversion of a $p\times p$ matrix and
evaluation of the log-likelihood over the full data set $\D$. This roughly yields
$\bigO{N p^{2+\epsilon}}$ running time complexity, which is prohibitive when $N$ and $p$ are large. In contrast, estimation with massive data sets typically requires a running time complexity
that is $\bigO{N p^{1-\epsilon}}$, i.e., that is linear in $N$ but
sublinear in the parameter dimension $p$.

Such performance is achieved in general by the
\gls{SGD} algorithm, which was initially proposed by \citet{sakrison1965} as a
modification of the Robbins-Monro procedure \citep{robbins1951} for recursive estimation. It is defined through the iteration
\begin{align}
\label{eq:explicit}
& \thetasgd{n} =  \thetasgd{n-1} + \gamma_n C_n \scoren{\thetasgd{n-1}}{n}.
\end{align}
We will refer to Equation \ref{eq:explicit} as \gls{SGD} with explicit updates, or
\emph{\gls{ESGD}} for short, because the next iterate $\thetasgd{n}$ can be computed immediately after the $n^{th}$ data point $(\xn, \yn)$ is observed.
The sequence $\gamma_n>0$ is the \emph{learning rate} sequence,
and is typically defined such that $n \gamma_n\to \gamma > 0$ as $n \to \infty$;
the hyperparameter $\gamma>0$ is fixed and known as the {\em learning rate parameter}.
The sequence $\{C_n\}$ is a sequence of positive-definite matrices, such that
$C_n \to C$ with $C$ known, and is used to better condition the iteration; in
the simplest case $C_n=I$, i.e., we simply use the identity matrix, which
results in {\em first-order} \gls{ESGD}.

From a computational perspective, \gls{ESGD} is efficient because
it replaces the expensive inversion of $p\times p$ matrices, as in
Fisher scoring, by a scalar sequence $\gamma_n >0$ and a matrix
$C_n$ that is fast to manipulate numerically, by design. Furthermore, the log-likelihood is evaluated at a single observation $\yn$ given $\xn$,
 rather than the entire data set $\D$, which saves
significant computation time.
From a theoretical perspective, \gls{ESGD} is justified because
the theory of stochastic approximations \citep[Theorem 1]{robbins1951}
implies that $\thetasgd{n}$ converges to a point
$\theta_\infty$ such that $\Ex{\scoren{\theta_\infty}{n}} =0$.
Under standard statistical theory, $\Ex{\scoren{\thetastar}{n}} =0$,
and this point is unique under typical regularity conditions \citep[Theorem 5.1, p.463]{lehmann1998theory}, such as concavity of log-likelihood; this is
true, for example, in the popular exponential family of statistical models
\citep{brown1986fundamentals}.
Therefore, $\theta_\infty=\thetastar$, i.e.,
\gls{ESGD} converges to the true parameter value.
In the finite $N$ setting, a similar condition holds where
$\thetasgd{n}$ approximates $\theta^{\mathrm{mle}}$ if the $n^{th}$
data point in Equation \ref{eq:explicit} is an unbiased sample from
the total $N$ data points; see also \citet{toulis2015scalable} for a
review of applications of \gls{SGD} on modern machine learning applications.

Despite these theoretical guarantees, \gls{ESGD} requires careful tuning of the
hyperparameter $\gamma$ in the learning rate: small values of the parameter make
the iteration \eqref{eq:explicit} very slow to converge in practice, whereas large values
can cause numerical divergence.
 Moreover, it is known that \gls{ESGD} is statistically inefficient even when $\gamma$ is correctly specified \citep{toulis2014statistical}.
In particular, the amount of information loss
from procedure \eqref{eq:explicit} depends on the spectral
gap of the Fisher information matrix, $\Fisher{\theta} = -\Ex{\nabla^2 \log f(\yn; \xn, \theta)}$,
calculated  at the true parameter value $\theta=\thetastar$. A large spectral
gap makes it hard, or even impossible, to make the learning rates
large enough for fast convergence, and also small enough for stability
\citep[Section 3.5]{toulis2015implicit}.

Motivated by these challenges, \citet{toulis2015stability} introduced \gls{AISGD},
which is defined by the procedure
\begin{align}
\label{eq:implicit}
 \thetaimpl{n} &=  \thetaimpl{n-1} + \gamma_n C_n \scoren{\thetaimpl{n}}{n},\\
 \label{eq:averaging}
 \thetaBar{n} & = (1/n) \sum_{i=1}^n \thetaimpl{i}.
\end{align}
The first key component of \gls{AISGD} is the \emph{implicit} update
\eqref{eq:implicit}. Note that it is implicit because the next iterate
$\thetaimpl{n}$ appears on both sides of the equation.
This simple modification of the \gls{ESGD} procedure offers several statistical
advantages.
In particular, assuming a common
starting point $\thetasgd{n-1} = \thetaimpl{n-1} \triangleq \vtheta_0$, one can show through a
Taylor approximation of \eqref{eq:implicit} around $\theta_0$ that the implicit
update satisfies
\begin{align}
\label{eq:implicit:shrinkage}
\Delta \thetaimpl{n} = (I + \gamma_n C_n \m{\hat{\mathcal{I}}} (\theta_0; \xn, \yn))^{-1} \Delta \thetasgd{n}
 + \bigO{\gamma_n^2},
\end{align}
where $\Delta \vthetan^{} = \vthetan - \vthetanprev$ for both methods, $I$
is the identity matrix, and
$\m{\hat{\mathcal{I}}} (\theta_0; \xn, \yn) = -\nabla^2 \ell(\vtheta_0;\xn,
\yn)$ is the observed Fisher information matrix at $\theta_0$ (equivalent
to the Hessian of the negative log-likelihood at $\theta_0$).
Equation \ref{eq:implicit:shrinkage} implies that the implicit update
\eqref{eq:implicit} is a  \emph{shrinked} version of the explicit update
\eqref{eq:explicit}. This shrinkage makes the iterations significantly more stable in
small-to-moderate samples, and also robust to misspecifications of the learning rate
parameter $\gamma$ \citep{toulis2014statistical}.
The implicit update \eqref{eq:implicit} also has a Bayesian interpretation, where
$\thetaim{n}$ is the posterior mode of a model with the standard multivariate
normal $\mathcal{N}(\thetaim{n-1}, \gamma_n  C_n)$ as the prior, and
$f(\theta; \xn, \yn)$ as the likelihood. Thus it provides an iterative form of
regularization.
In optimization, update \eqref{eq:implicit} is known as a \emph{proximal update},
and corresponds to a stochastic version of the proximal point algorithm
\citep{rockafellar1976monotone}.
\citet{krakowskigeometric} and \citet{nemirovski2009robust} have shown that
proximal methods fit better in the geometry of the parameter space.

The second key component of \gls{AISGD} is iterate averaging
\eqref{eq:averaging}, which guarantees optimal statistical efficiency under fairly
relaxed conditions. \citet{ruppert1988efficient} and \citet{polyak1992acceleration}
first proved that averaging of iterates can achieve statistical optimality
in the standard context of stochastic approximation with explicit updates;
\citet{toulis2015stability} extended this result to the \gls{ISGD} update \eqref{eq:implicit}.
Thus, \gls{AISGD} is effectively a recursive estimation method that is
both statistically optimal and numerically stable, while remaining applicable to the setting of massive and/or streaming data.

In this paper we develop statistically efficient \gls{SGD} algorithms for
generalized linear models---extending Algorithm 1 of \citet{toulis2014statistical}---and also develop \gls{SGD} algorithms to perform high-dimensional M-estimation.
This allows for scalable estimation of such models with massive and/or
streaming data. We provide a publicly available package \pkg{sgd}
\citep{tran2015sgd} written in \proglang{R}, which implements \gls{AISGD}, as
well as other \gls{SGD} variants.
In Section \ref{sec:algorithms}, we develop the algorithms.
Section \ref{sec:experiments} contains experiments on simulated
and real-world data, in which we demonstrate
the advantages of the \pkg{sgd} package compared to alternative software.
In Section \ref{sec:interface}, we describe the interface of \pkg{sgd} and implementation details for its use in practice.

\section{Algorithms}
\label{sec:algorithms}
In this section we develop algorithms which implement \gls{ISGD} and
\gls{AISGD} for generalized linear models as well as M-estimation.
We start by introducing an algorithm which efficiently computes a generalization
of implicit update \eqref{eq:implicit}, which is useful for the aforementioned
applications.

\subsection{Efficient computation of implicit updates}
\label{sec:algorithms:efficient}
The main difficulty in applying \gls{AISGD} is the solution of the
multidimensional fixed point equation for the implicit update
\eqref{eq:implicit}. In the large class of models where the likelihood given
covariate $\x$ depends
on the parameter $\theta$ only through the {\em natural parameter}
$\eta\equiv \x^\top \theta$, the solution of the fixed-point equation is
computationally efficient.
The general result is given in Theorem \ref{theorem:algo:implicit},
whereas the assumption is made more precise below.

\begin{assumption}
\label{assumption:linear}
The likelihood $\ell(\theta; \xn, \yn)
 \equiv \log f(\yn; \xn, \theta)$ of parameter value $\theta$ given data point $(\xn, \yn)$
depends on $\theta$ only through the product $\xn^\top \theta$, i.e.,
\begin{align}
\ell(\theta; \xn, \yn) \equiv \ell(\xn^\top \theta; \xn, \yn).
\end{align}
\end{assumption}
\newcommand{\LinearAssumption}{Assumption \ref{assumption:linear}\xspace}
A key implication of \LinearAssumption is that the direction of the gradient of the log-likelihood does {\em not} depend on the parameter value since
$\nabla \log f(\yn; \xn, \theta) = \ell'(\xn^\top\theta; \xn, \yn) \xn$,
where the latter derivative is with respect to the natural parameter $\xn^\top\theta$
and with fixed data $\xn, \yn$. This property is crucial
because it implies that the implicit update \eqref{eq:implicit} can be
performed once a scalar value is found that will
appropriately scale the gradient.

\vspace{10px}
\newcommand{\LinearTheorem}{
Suppose \LinearAssumption holds.
Then the gradient for the implicit iterate $\thetaim{n}$
\eqref{eq:implicit} is a scaled version of
the gradient at the previous iterate, i.e.,
\begin{align}
\label{algo:implicit:scale}
\nabla \log f(\yn; \xn, \thetaim{n}) =
   s_n \nabla \log f(\yn; \xn, \thetaim{n-1}).
\end{align}
The scalar $s_n\in\mathbb{R}$ satisfies
\begin{align}
\label{algo:implicit:fp}
s_n \kappa_{n-1} =
 \ell'\left(\xn^\top \thetaim{n-1} + \gamma_n s_n \kappa_{n-1} \xn^\top C_n \xn;  \xn, \yn \right ),
\end{align}
where $\kappa_{n-1} = \ell'(\xn^\top \thetaim{n-1}; \xn, \yn)$.
}
\begin{theorem}
\label{theorem:algo:implicit}
\LinearTheorem
\end{theorem}

Theorem \ref{theorem:algo:implicit} shows that the gradient $\scoren{\thetaimpl{n}}{n}$
in the implicit update \eqref{eq:implicit} is in fact a scaled version of the
gradient $\scoren{\thetaimpl{n-1}}{n}$ that would appear in update \eqref{eq:implicit} if we were applying
explicit updates. Therefore, computing the implicit update reduces to finding
the scale factor $s_n\in\mathbb{R}$. See \citet[Threorem 4.1]{toulis2015implicit} for a proof.
%
%

\parhead{Penalized likelihood.}
It is possible to regularize both explicit and implicit \gls{SGD} by adding a penalty
to the log-likelihood. In particular, we consider the elastic net \citep{zou2005regularization}, where
for some fixed $\alpha\in[0,1]$ the penalty function is
\begin{equation}
P_\alpha(\theta) = (1-\alpha)\frac{1}{2}\|\theta\|_2^2 + \alpha\|\theta\|_1.
\end{equation}
Adding the elastic net with a regularization parameter $\lambda\in \Reals{}$
to \gls{ESGD} is straightforward:
\begin{align}
\thetasgd{n} = \thetasgd{n-1} + \gamma_n C_n (\nabla \log f(\yn; \xn, \thetasgd{n-1}) -
\lambda \nabla P_{\alpha}(\thetasgd{n-1})),
\end{align}
where the gradient of the elastic net penalty is given by
\begin{equation}
\nabla P_\alpha(\thetaim{n-1}) =
(1-\alpha)\thetasgd{n-1} +
\alpha\operatorname{sign}(\thetasgd{n-1}).
\end{equation}
Here, the operation $\operatorname{sign}(\theta)$ is the element-wise
sign operation, outputting 1 if $\theta_j>0$, $-1$ if $\theta_j<0$, and 0 otherwise.

For \gls{ISGD} the update would be
\begin{align}
\label{eq:penalized:im_bad}
\thetaim{n} = \thetaim{n-1} + \gamma_nC_n  (\nabla \log f(\yn; \xn, \thetaim{n}) -
\lambda \nabla P_{\alpha}(\thetaim{n})).
\end{align}
However, it is not generally possible to compute update  \eqref{eq:penalized:im_bad}. For example, \LinearAssumption\ does
not hold because the gradient of the log-likelihood and the gradient of the
penalty generally have two different directions.
This breaks the argument of Theorem \ref{theorem:algo:implicit},
where the direction of the update calculated at the next iterate $\thetaim{n}$
is the same as the direction of the update calculated at
the previous iterate $\thetaim{n-1}$.

To circumvent this problem, we simply penalize the previous iterate instead
of the current, i.e., perform the update
\begin{align}
\label{eq:penalized:im_good}
\thetaim{n} = \thetaim{n-1} + \gamma_n C_n (\nabla \log f(\yn; \xn, \thetaim{n}) -
\lambda \nabla P_{\alpha}(\thetaim{n-1})).
\end{align}
Then update \eqref{eq:penalized:im_good} is equivalent to
\begin{align}
\label{eq:penalized:im_good2}
\thetaim{n} = \thetaim{n-1} + \gamma_n C_n (s_n \nabla \log f(\yn; \xn, \thetaim{n-1}) -
\lambda \nabla P_{\alpha}(\thetaim{n-1})),
\end{align}
where the scale factor $s_n$ satisfies
\begin{align}
\label{eq:implicit:penalty}
s_n \kappa_{n-1} =
 \ell'\left(\xn^\top \thetaim{n-1} - \gamma_n \lambda\xn^\top C_n \nabla P_{\alpha}(\thetaim{n-1}) + \gamma_n s_n \kappa_{n-1} \xn^\top C_n \xn  ;  \xn, \yn \right ),
\end{align}
and where $\kappa_{n-1} = \ell'(\xn^\top\thetaim{n-1}; \xn, \yn)$.
A proof for this case with penalized likelihoods is identical to the proof of
Theorem \ref{theorem:algo:implicit}.

\parhead{Final algorithm for implicit updates.}
This analysis leads to Algorithm \ref{algo:implicit}, which, for models satisfying
\LinearAssumption, implements
the most general update \eqref{eq:penalized:im_good2} of \gls{ISGD} with
conditioning matrices and penalty.  This algorithm applies a root-finding procedure solving
Equation \ref{eq:implicit:penalty} at every iteration, which
is fast because the equation is one-dimensional and the search bounds for
the solution are known, having a diminishing range $\bigO{\gamma_n}$.
Indeed, the one-dimensional search is computationally negligible in practice, as we see in Section \ref{sec:experiments}.

We also note that because the implicit update \eqref{eq:implicit:glm} effectively
does regularization as a shrinkage estimate (see Equation \ref{eq:implicit:shrinkage}),
the use of penalization is
not as crucial in practice as it is for explicit updates. We make extensive experiments
using Algorithm \ref{algo:implicit_glm} and also examine this effect in Section \ref{sec:experiments}.

\begin{algorithm}[t]
\caption{Efficient implementation of implicit update \eqref{eq:penalized:im_good2}}
\begin{algorithmic}[1]
  \Function{implicit\_update}{$\ell'(\cdot;\cdot),\gamma_n,\thetaim{n-1},\xn,\yn,C_n, P_{\alpha}$}
    \State \emph{\small \# Compute search bounds} $B$
    \State $r_n \gets \gamma_n \ell'\left(\xn^\top \thetaim{n-1}; \xn,\yn\right )$
    \State $B \gets [0, r_n]$
    \If{$r_n \le 0$}
      \State $B \gets [r_n, 0]$
    \EndIf
    \State  \emph{\small \# Solve fixed-point equation by a
    root-finding method}
    \State $\xi =  \gamma_n \ell'(\xn^\top \thetaim{n-1}- \gamma_n \lambda\xn^\top C_n \nabla P_{\alpha}(\thetaim{n-1}) + \xi \xn^\top C_n \xn;
    \xn,\yn)$, $\xi\in B$
    \State $s_n \gets \xi/r_n$
    \State \emph{\small \# Equivalent to implicit update \eqref{eq:penalized:im_good2}}
    \State \Return $\thetaim{n-1} + \gamma_n C_n
	\left(s_n \ell'\left(\xn^\top  \thetaim{n-1}; \xn, \yn\right )\xn - \lambda \nabla P_{\alpha}(\thetaim{n-1})\right)$
 \EndFunction
\end{algorithmic}
\label{algo:implicit}
\end{algorithm}

\subsection{Generalized linear models}
\label{sec:algorithms:glm}
In the family of \glspl{GLM}, the outcome $y_n \in
\Reals{}$  follows an exponential family distribution conditional on
$\xn$,
\begin{equation}
y_n \mid \xn \sim \exp\left\{\frac{1}{\psi} (\eta_ny_n - b(\eta_n))\right\}c(y_n,\psi),
\quad
\eta_n\equiv \xn^\top\thetastar,
\end{equation}
where the scalar $\psi>0$ is the dispersion parameter which affects the
variance of the outcome, $c(\cdot, \cdot)$ is the base measure, and
$b(\cdot)$ is the log normalizer which ensures that the distribution integrates
to one.\footnote{We present one-dimensional outcomes for simplicity.
However, our theory easily extends to multidimensional outcomes. Such an
extension is given, for example, in Section \ref{sec:algorithms:m} on M-estimation.}
Additionally, in a \gls{GLM} it is assumed that $\ExCond{y_n}{\xn} = h(\xn^\top
\thetastar)$, where $h : \Reals{}\to\Reals{}$ is known as the \emph{transfer function}
\citep{nelder1972, dobson2008introduction}.
A simple property of \glspl{GLM} is that the
transfer function is the first derivative of the log normalizer, i.e.,
$h(\xn^\top\theta) = b'(\xn^\top\theta)$, for all $\xn, \theta$.

A straightforward implementation of \gls{ESGD} for estimation with
\glspl{GLM} is
\begin{equation}
\label{def:sgd:explicit_glm}
\thetasgd{n} = \thetasgd{n-1} + \gamma_n C_n  [y_n - h(\xn^\top
\thetasgd{n-1})] \xn.
\end{equation}
Similarly, the \gls{AISGD} procedure can be written as
\begin{align}
\label{eq:implicit:glm}
\begin{split}
\thetaim{n} & = \thetaim{n-1} + \gamma_n C_n  [y_n - h(\xn^\top \thetaim{n})]
\xn,\\
\thetaBar{n} &= \frac{1}{n} \sum_{i=1}^n \thetaim{n}.
\end{split}
\end{align}
By assumption, $\ell(\theta; y_n, \xn) \propto (\xn^\top\thetastar)y_n - b(\xn^\top\thetastar)$, and thus the log-likelihood depends on parameter value $\thetastar$ only through its linear combination with covariate value $\xn$.
Additionally, $\Var{y_n|\xn} = h'(\xn^\top\thetastar) ||\xn||^2$,
and thus $h' \ge 0$, which implies that $\ell$ is twice-differentiable and concave,
thus fulfilling \LinearAssumption.

\textbf{Penalized likelihood.}
As argued before, one can add the elastic penalty by applying it to the previous estimate instead of the current.
That is, for fixed $\alpha\in[0,1]$ and regularization parameter
$\lambda\in\mathbb{R}$, the \gls{AISGD} procedure for generalized linear models
with elastic net is
\begin{align}
\begin{split}
\thetaim{n} & = \thetaim{n-1} + \gamma_n C_n \left([y_n - h(\xn^\top \thetaim{n})]
\xn -
\lambda \nabla P_\alpha(\thetaim{n-1})\right),\\
\thetaBar{n} &= \frac{1}{n} \sum_{i=1}^n \thetaim{n}.
\end{split}
\label{eq:implicit:glm:elastic}
\end{align}
Algorithm \ref{algo:implicit_glm} implements estimation of GLMs through \gls{AISGD} based
on updates \eqref{eq:implicit:glm:elastic}.
\begin{algorithm}[t!]
\begin{algorithmic}[1]
\caption{Estimation of \acrlongpl{GLM} with \gls{AISGD}}
\label{algo:implicit_glm}
  \State Initialize $\thetaim{0}, \thetaBar{0}$
  \For{$n=1,2,\ldots$}
\State Define $\ell'(\xn^\top\theta; \xn, \yn) \equiv
	y_n - h(\xn^\top \theta)$
\State Calculate implicit update
\begin{equation*}
\thetaim{n} \gets \footnotesize{\textsc{IMPLICIT\_UPDATE}}(\ell'(\cdot;\cdot),\gamma_n,\thetaim{n-1},\xn,\yn,C_n, P_{\alpha})
\end{equation*}
   \State $\thetaBar{n} \gets \frac{n-1}{n}\thetaBar{n-1} +
        \frac{1}{n}\thetaim{n}$
  \EndFor
\end{algorithmic}
\end{algorithm}

\subsection{M-Estimation}
\label{sec:algorithms:m}
Given a data set of $N$ observations $\mathcal{D}=\{(\xn, \yn)\}$ and a convex function $\rho :
\Reals{} \to \Reals{+}$, the \gls{M} is defined as
\begin{align}
\label{def:m}
\hat{\theta}^m = \arg \min_{\theta\in\Reals{p}} \sum_{n=1}^N \rho(\yn - \xn^\top \theta),
\end{align}
where it is assumed $\yn = \xn^\top \thetastar + \epsilon_n$, and $\epsilon_n$
are i.i.d. zero mean-valued noise.
\glspl{M} are especially useful in robust
statistics \citep{huber1964robust, huber2009robust}, as appropriate choice
of $\rho$ can reduce the influence of outliers in data.
Recently, there has been increased interest in the
literature for fast approximation of \glspl{M} due to their robustness \citep{donoho2013high, jain2014iterative}.

Typically in \glslink{M}{M-estimation}, $\rho$ is twice-differentiable around
zero and
\begin{align}
\Ex{\rho'(\yn - \xn^\top \hat{\theta}^m) \xn} = 0,
\end{align}
where the expectation is over the empirical data distribution.
Therefore \gls{SGD} algorithms can be applied to approximate the \gls{M}
$\hat{\theta}^m$.
Importantly, $\rho$ is convex, which implies that the
conditions of \LinearAssumption are met.

The \gls{AISGD} procedure for approximating \glspl{M} is
\begin{align}
\label{eq:implicit:m}
\thetaim{n} & = \thetaim{n-1} + \gamma_n C_n  [\rho'(\yn -
\xn^\top \thetaim{n})]
\xn,\\
\thetaBar{n} &= \frac{1}{n} \sum_{i=1}^n \thetaim{n}.
\end{align}
An outline of the procedure is given in Algorithm \ref{algo:implicit_m}.
As before, Algorithm \ref{algo:implicit_m} also includes the optional use of a sequence of conditioning matrices $C_n$ and a
penalty function $P_\alpha$. The use of penalization has particularly been
considered as a way to merge the robustness properties given by a choice of
$\rho$ with sparsity, e.g, through lasso
\citep{owen2007robust,lambert2011robust,li2011nonconcave}.
\begin{algorithm}[t]
\begin{algorithmic}[1]
  \caption{\glslink{M}{M-estimation} with \gls{AISGD}}
  \label{algo:implicit_m}
  \State Initialize $\thetaim{0}, \thetaBar{0}$
  \For{$n=1,2,\ldots$}
  \State Define $\ell'(\xn^\top\theta; \xn, \yn) \equiv
	 -\rho'(\yn-\xn^\top \theta)$
\State Calculate implicit update
\begin{equation*}
\thetaim{n} \gets \footnotesize{\textsc{IMPLICIT\_UPDATE}}(\ell'(\cdot;\cdot),\gamma_n,\thetaim{n-1},\xn,\yn,C_n, P_{\alpha})
\end{equation*}
   \State $\thetaBar{n} \gets \frac{n-1}{n}\thetaBar{n-1} +
        \frac{1}{n}\thetaim{n}$
  \EndFor
\end{algorithmic}
\end{algorithm}

It is also typical to assume that the density of $\epsilon_n$ is
symmetric around zero. Therefore, it also holds $\Ex{\rho'(\yn - \xn^\top
\thetastar) \xn} = 0$, where the expectation is over the true data distribution.
Hence \gls{SGD} procedures can be used to estimate $\thetastar$ in the case
of an infinite stream of observations ($N = \infty$). We write
Algorithm \ref{algo:implicit_m} for the case of finite $N$, but it is trivial to adapt
the procedure to infinite $N$.

\section{Experiments}
\label{sec:experiments}
In this section, we compare the \gls{SGD} methods implemented
in the \pkg{sgd} package,
such as \gls{ESGD} and \gls{AISGD}, with standard, deterministic
optimization methods that are widely used in statistical practice,
such as \pkg{glmnet}, \pkg{biglm}, and \pkg{speedglm}.
We demonstrate in both massive and streaming data settings
that standard methods are not applicable, and furthermore that \gls{SGD} methods
outperform such methods upon orders of magnitude in runtime and convergence.

As standard methods are not competitive, we also compare the
proposed \gls{SGD} methods to each other, e.g., comparing
\gls{AISGD} to \gls{ESGD}, across a wide range of learning rate specifications,
including adaptive specifications such as AdaGrad \citep{duchi2011adaptive}
and RMSProp \citep{tieleman2012lecture}; more details on the
specifications which are available in \pkg{sgd} are given in Section \ref{sec:interface:learning}.

All timings are carried out on a general-purpose 2.6 GHz Intel Core i5 processor, and are reported for various algorithms which reach a thresholded $L_2$ distance to the true parameter value.

\subsection{Linear regression with the lasso}
\label{sec:experiments:lasso}
We follow an experiment used in benchmarking the \pkg{glmnet} package
\citep[Section 5.1]{friedman2010regularization}, which fits \glspl{GLM} with the
elastic net penalty over a regularization path.
As \pkg{glmnet} was shown to outperform related software such as
\pkg{elasticnet} \citep{zou2012elasticnet} and \pkg{lars}
\citep{hastie2013lars}, we compare \pkg{sgd} strictly to \pkg{glmnet}.
The design matrix $\X$ with $N$ observations and $p$ predictors is generated from
a normal distribution such that each pair of predictors $\X_j,\X_{j'}$ has the same
correlation $\rho$.
Each of the $N$ outcomes $\yn$, $n=1, 2, \ldots N$, is
defined as
\begin{equation}
\yn = \xn^\top \thetastar  + k\epsilon_n,
\end{equation}
where ${\thetastar}_j = (-1)^j\exp(-2(j-1)/20)$ so that
the elements of the true parameter value $\thetastar$ have alternating signs and
are exponentially decreasing.  The noise $\epsilon_n$ is distributed as a
standard normal, $\epsilon\sim\mathcal{N}(0,1)$,
and $k$ is chosen so that the signal-to-noise ratio is equal to 3.0.
We run \pkg{glmnet} with ``covariance updates", which takes advantage of sparse
updates in the parameter space to reduce the complexity of
$\mathcal{O}(Np)$ calculations per iteration. It performs better in
our experiments than the ``naive update" also considered in
\citet{friedman2010regularization}.

\begin{table}[!t]
  \begin{center}
  \begin{tabular}{lllllll}
  \cmidrule{2-7}
  &\multicolumn{6}{c}{Correlation}\\
  & 0 & 0.1 & 0.2 & 0.5 & 0.9 & 0.95 \\
  \cmidrule{2-7}\\
  &\multicolumn{6}{c}{$N=1,000$\, $p=100$ (sec)}\\
  \cmidrule{2-7}
  \code{sgd(method="ai-sgd")} & 0.03 & 0.03 & 0.03 & 0.03 & 0.04 & 0.34\\
  \code{sgd(method="sgd")} & 0.02 & 0.02 & 0.02 & 0.02 & 0.03 & 0.03\\
  \code{glmnet} & 0.02 & 0.02 & 0.02 & 0.02 & 0.02 & 0.03\\
  \cmidrule{2-7}\\
  &\multicolumn{6}{c}{$N=10,000$\, $p=1,000$ (sec)}\\
  \cmidrule{2-7}
  \code{sgd(method="ai-sgd")} & 1.81 & 1.65 & 1.78 & 1.50 & 1.85 & 1.83\\
  \code{sgd(method="sgd")} & 2.78 & 2.90 & 2.93 & 2.81 & -- & --\\
  \code{glmnet} & 6.60 & 7.76 & 8.00 & 7.83 & 6.50 & 6.70\\
  \cmidrule{2-7}
  &\multicolumn{6}{c}{$N=50,000$\, $p=10,000$ (min)}\\
  \cmidrule{2-7}
  \code{sgd(method="ai-sgd")} & 3.12 & 3.51 & 3.43 & 3.26 & 3.40 & 3.38\\
  \code{sgd(method="sgd")} & 4.83 & 4.86 & 5.23 & -- & -- & --\\
  \code{glmnet} & 14.58 & 15.28 & 16.29 & 15.58 & 16.54 & 16.41\\
  \cmidrule{2-7}
  &\multicolumn{6}{c}{$N=1,000,000$\, $p=50,000$ (min)}\\
  \cmidrule{2-7}
  \code{sgd(method="ai-sgd")} & 22.23 & 21.10 & 19.88 & 21.52 & 18.53 & 20.53\\
  \code{sgd(method="sgd")} & 27.80 & 34.08 & -- & -- & -- & --\\
  \code{glmnet} & -- & -- & -- & -- & -- & --\\
  \cmidrule{2-7}
  &\multicolumn{6}{c}{$N=10,000,000$\, $p=100,000$ (hr)}\\
  \cmidrule{2-7}
  \code{sgd(method="ai-sgd")} & 9.38 & 10.20 & 9.58 & 8.54 & 10.11 & 10.74\\
  \code{sgd(method="sgd")} & 13.50 & -- & -- & -- & -- & --\\
  \code{glmnet} & -- & -- & -- & -- & -- & --\\
  \cmidrule{2-7}
  \end{tabular}
  \end{center}
  \caption{Linear regression with the lasso. Timing (in various units) is
  displayed for 100 $\lambda$ values, averaged over 10 runs. The first line is
  \pkg{sgd} using \gls{AISGD} and the second line is \pkg{sgd} using \gls{ESGD}.
  Omitted entries indicate failure of the algorithm; for \gls{ESGD} it
  numerically diverges, and for \pkg{glmnet} it could not run due to memory
  limitations.
  }
  \label{table:lasso}
\end{table}
Table \ref{table:lasso} outlines results for a combination of triplets $(N,p,\rho)$,
ranging from $N=1,000$ observations and scaling up to $N=10$ million. \pkg{glmnet} is seen to be competitive with \gls{SGD} procedures under the
setting of $N=1,000$ observations, and in fact \pkg{glmnet} slightly outperforms
\gls{SGD} algorithms for lower dimensions of $N$ and $p$. It
is in any higher dimensional setting where \pkg{sgd} strictly dominates
\pkg{glmnet}, as seen in the table where for example, with $N=50,000$ and $p=10,000$,
\pkg{sgd} is orders of magnitude faster.

Furthermore, \pkg{glmnet} is restricted by the memory limitations of computer
hardware. For example, simulations with $100,000$ observations and $10,000$
features require 8 GB in memory for simply storing the data, and more is
required for parameter storage and computational overhead. For the \pkg{sgd} package, we simply stream the data points using
\pkg{bigmemory},
which requires less than 500 MB of RAM for all our experiments, a 16-fold decrease
in memory requirements. This is not
possible for \pkg{glmnet} in either the case of real streaming data, or simply as
a way to remove memory bottlenecks. In principle, gradient descent algorithms
such as \pkg{glmnet} can read and destroy data memory from disk as it
loops over the full data set; however, this is impractical as it requires such
an expensive memory access at each iteration.

We now compare the \gls{SGD} algorithms. For small dimensional problems,
\gls{ESGD} achieves faster runtime than \gls{AISGD} as it does not require a
one-dimensional search following Algorithm \ref{algo:implicit}. However, in high
dimensions and high correlations, it becomes extremely difficult for \gls{ESGD}
to even converge for this toy linear model. It is sensitive to the learning rate, and any misspecification can cause it to diverge numerically.
Thus, we were not able to obtain a proper timing for \gls{ESGD} in settings of
either high correlation ($\rho > 0.9$) or high dimension with medium correlation
($\rho > 0.5$).
In practice one must tune the hyperparameter for
\gls{ESGD}---thus requiring significant computational overhead and user
input---while also closely monitoring the stochastic gradients for consideration
of other numerical issues. \gls{AISGD} on the other hand uses additional
computation per iteration, which in high dimensions is negligible compared to the
cost of a stochastic gradient update. This additional computation leads to
significantly more robust updates and faster convergence.

\subsection{Logistic regression with ridge penalty}
\label{sec:experiments:logistic}
\begin{figure}[!p]
  \begin{minipage}[b]{\linewidth}
  \centering
  \begin{tabular}{llllllll}
  \toprule
  data set & \pkg{sgd} (\gls{AISGD}) & \pkg{sgd} (\gls{SGD}) & \pkg{biglm} &
  \pkg{speedglm} & \pkg{LiblineaR} & \pkg{mnlogit} & \code{glm.fit}\\
  \midrule
  covtype & 5.21 & 7.58 & -- & -- & 1444.78 & 16.04 & 40.11\\
  delta & 10.10 & 10.23 & 736.13 & 30.50 & 2167.14 & 445.73 & 498.97\\
  rcv1 & 14.15 & 15.42 & -- & -- & 133.10 & -- & --\\
  mnist & 3.50 & 3.37 & -- & -- & 208.55 & 232.53 & 890.76\\
  \bottomrule
  \end{tabular}
  \captionof{table}{Large-scale logistic regression on four data sets. Timing
  (in seconds) is displayed, averaged over 10 runs. Omitted entries indicate
  failure of the algorithm; for \pkg{biglm} and \pkg{speedglm}, it could not run
  due to inversions of singular matrices; for \pkg{mnlogit} it could not run
  due to memory limitations.}
  \label{table:logistic}
  \end{minipage}
  \\
  \begin{minipage}[b]{0.5\linewidth}
    \centering
    \includegraphics[width=0.95\textwidth]{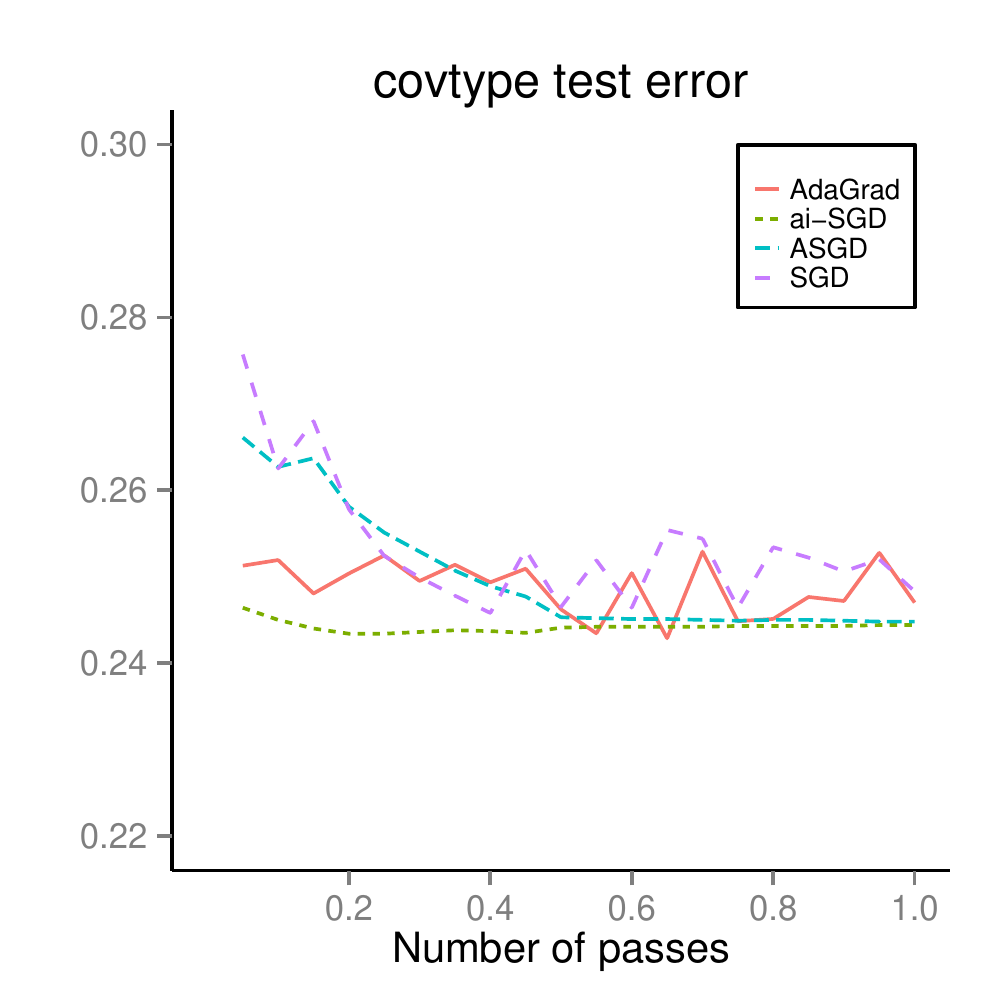}
    \phantomsubcaption \label{subfig:1}
  \end{minipage}
  \hfill
  \begin{minipage}[b]{0.5\linewidth}
    \centering
    \includegraphics[width=0.95\textwidth]{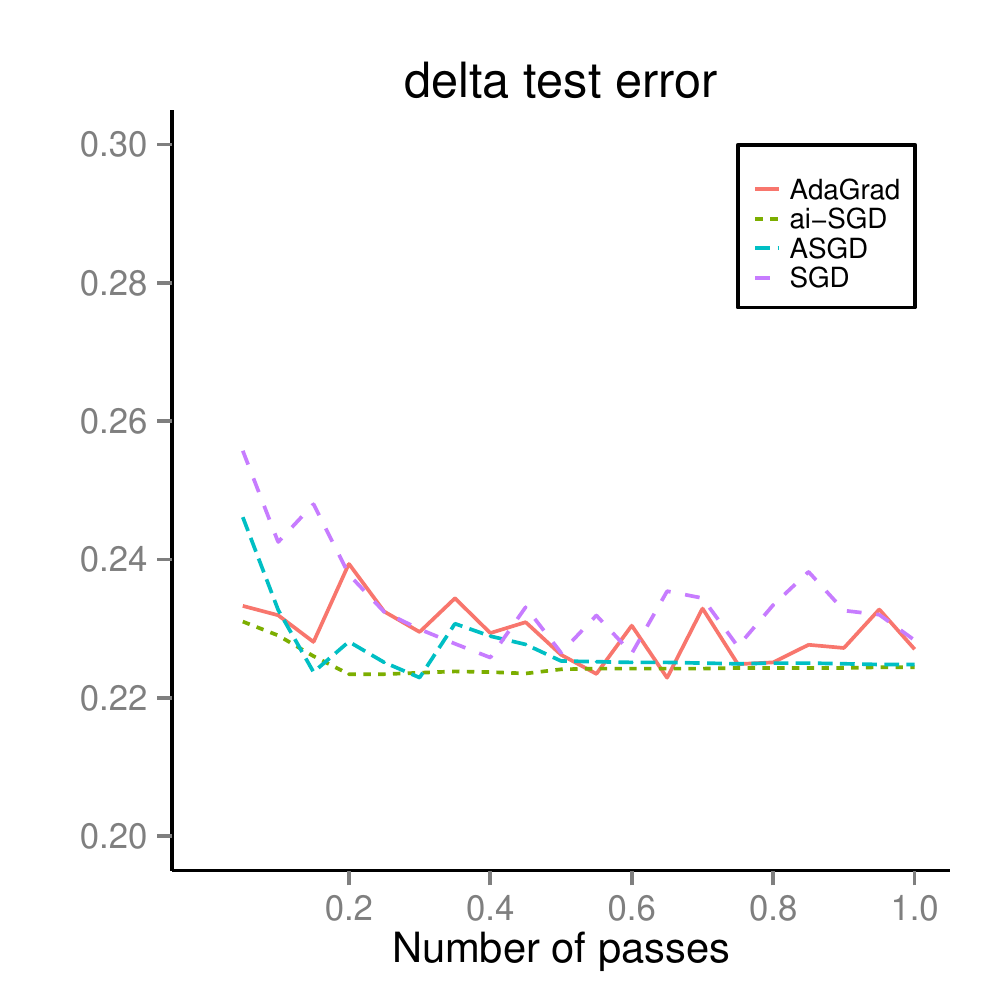}
    \phantomsubcaption \label{subfig:2}
  \end{minipage}
  \\
  \begin{minipage}[b]{0.5\linewidth}
    \centering
    \includegraphics[width=0.95\textwidth]{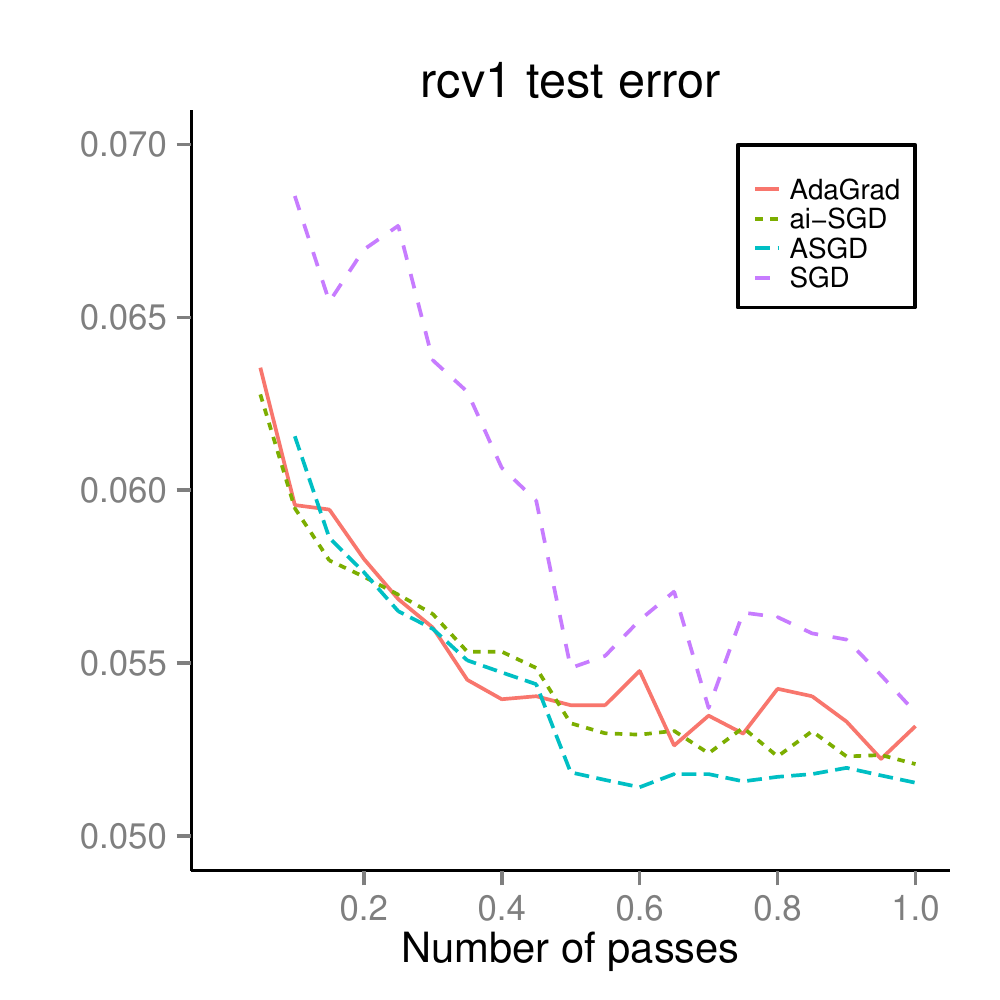}
    \phantomsubcaption \label{subfig:3}
  \end{minipage}
  \hfill
  \begin{minipage}[b]{0.5\linewidth}
    \centering
    \includegraphics[width=0.95\textwidth]{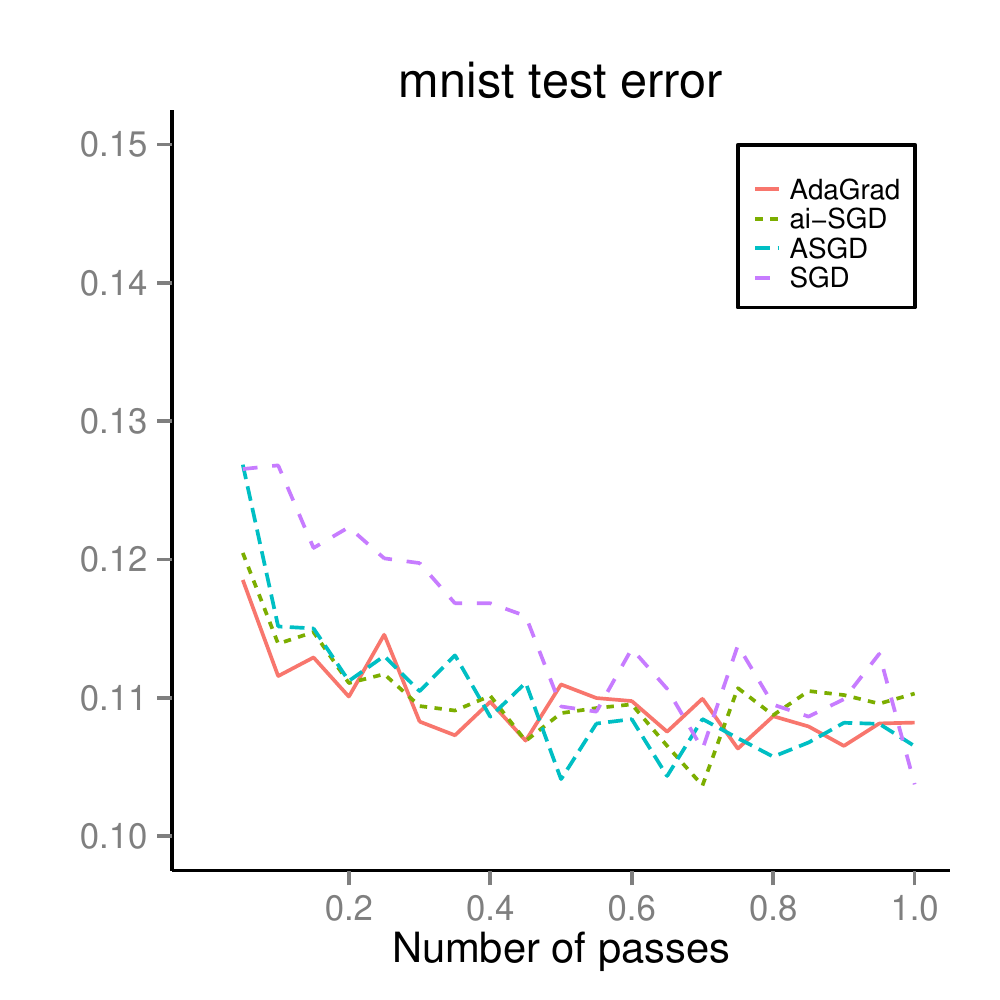}
    \phantomsubcaption \label{subfig:4}
  \end{minipage}
  \caption{Large scale logistic regression on four data sets. Each plot
  indicates the classification error on the test set for \gls{ESGD} with
  AdaGrad, \gls{AISGD}, \gls{ASGD}, and \gls{ESGD} over a pass of the data.}
  \label{figure:logistic}
\end{figure}

Following benchmarks that are popular in the machine learning and optimization
literature \citep{xu2011towards, shamir2012stochastic, bach2013non,
schmidt2013minimizing}, we perform large-scale logistic regression on four data
sets:
\begin{itemize}
\item
\code{rcv1} \citep{lewis2004rcv1}: text data set in which the task is to
classify documents belonging to class \code{ccat}, where we apply
preprocessing provided by \citet{bottou2012stochastic}.
\item
\code{covtype} \citep{blackard1998comparison}: data set consisting of forest
cover types in which the task is to classify for one specific class among 7 forest cover types.
\item
\code{delta} \citep{sonnenburg2008pascal}: synthetic data offered in the PASCAL
Large Scale Challenge. We apply the default processing offered by the
challenge organizers.
\item
\code{mnist} \citep{lecun1998gradient}: images of handwritten digits, where the
task is to classify digit 9 against all others.
\end{itemize}
A summary of the data sets is available in Table \ref{table:logistic}, where the
number of observations are typically on the order of several hundred thousand,
and the covariates range from a few dozen to tens of thousands. The
regularization parameter $\lambda$ for the ridge penalty are set according to
those used in \citet{xu2011towards}.

\begin{table}[tb]
  \centering
  \begin{tabular}{lllllllllll}
  \toprule
         & description          & type   & covariates& training set & test set & $\lambda$\\
  \midrule
  covtype& forest cover type    & sparse & 54      & 464,809 & 116,203 & $10^{-6}$\\
  delta  & synthetic data       & dense  & 500     & 450,000 & 50,000     & $10^{-2}$\\
  rcv1   & text data            & sparse & 47,152  & 781,265 & 23,149  & $10^{-5}$\\
  mnist  & digit image features & dense  & 784     & 60,000  & 10,000  & $10^{-3}$\\
  \bottomrule
  \end{tabular}
  \caption{Summary of data sets and the $L_2$ regularization parameter
  $\lambda$ used.}
  \label{table:data}
\end{table}

We compare to the following three packages: \pkg{biglm} \citep{lumley2013biglm}
and \pkg{speedglm} \citep{enea2015speedglm}, both of which perform approximate
updates using iteratively reweighted least squares, and \pkg{LiblineaR}
\citep{helleputte2015liblinear}, which is a simple wrapper to a \proglang{C++}
library for regularized linear classification. We use the stochastic dual
coordinate ascent algorithm \citep{shalev2013stochastic} in \pkg{LiblineaR}.
In addition, we consider the \pkg{mnlogit} package \citep{hasan2015mnlogit},
which implements multinomial logistic regression using the classical technique
of Newton-Raphson, and exploits iterations over intermediate data structures for
fast Hessian calculations. For modest-sized problems, \pkg{mnlogit} is shown to
be 10-50 times faster than \pkg{mlogit} \citep{croissant2013mlogit}, \pkg{VGAM}
\citep{yee2010vgam}, and the \code{multinom} function in \pkg{nnet}
\citep{venables2002modern}. Finally, we also run the default function
\code{glm.fit} as a baseline. We note that \pkg{mnlogit} and
\code{glm.fit} can be only employed for standard (unregularized) multinomial
regression, so we run them without the ridge penalty.

Table \ref{table:logistic} outlines the runtimes for the considered packages. The
two \gls{SGD} algorithms are orders of magnitude faster than its competitors on
all data sets. Interestingly, \pkg{biglm} and \pkg{speedglm} failed to run on
the three real data sets when attempting to invert subsets of the data, and
only succeeded for the one synthetic data set \code{delta}. We also note that
the largest data set---\code{rcv1}---failed for the majority of algorithms:
only the packages \pkg{sgd} and \pkg{LiblineaR} were able to converge, both of
which natively use stochastic gradients for computationally efficient updates.
However, \pkg{sgd} is significantly faster
because less overhead seems to be involved in
passing data structures to perform computation in native \proglang{C++}.

Moreover, \pkg{sgd} requires $\mathcal{O}(p)$ memory, which is
optimal in the sense that $\mathcal{O}(p)$ is the minimum required for simply
storing the $n^{th}$ iterate $\thetaim{n}$. Both \pkg{biglm} and \pkg{speedglm} require
$\mathcal{O}(p^2)$ for the inversion of a $p\times p$ matrix, as do
\pkg{mnlogit} and \code{glm.fit}.
The \pkg{mnlogit} package also requires data in the \code{long} format, which
leads to a duplication of rows, as many entries display redundant information.
Moreover, while exploitation of the Hessian structure
can help in practice (as it outperforms \code{glm.fit}), we observe that the
traditional technique of Newton-Raphson remains untenable because it still requires $\mathcal{O}(Np^2)$ complexity per
iteration in the worst case.

For demonstration, Figure \ref{figure:logistic} shows the progress of multiple
\gls{SGD} algorithms available in \pkg{sgd} (see
Section \ref{sec:interface:stochastic}) over a pass of the data. We note that
\gls{AISGD} achieves the fastest or competitive convergence rates, without
requiring significant tuning of parameters as the other algorithms do; this
includes popular adaptive learning rate specifications, such as \gls{ESGD}
with AdaGrad.

\subsection{M-estimation with the Huber loss}
\label{sec:experiments:huber}
We follow an example for high-dimensional M-estimation in
\citet[Section 2.4]{donoho2013high}.
Define the convex function $\rho:\mathbb{R}\to\mathbb{R}^+$ to be the \emph{Huber loss},
\begin{align*}
\rho(z; \lambda) = \begin{cases}
z^2/2, &\text{ if }|z|\le\lambda,\\
\lambda|z| - \lambda^2/2, &\text{otherwise}.
\end{cases}
\end{align*}
Fix the thresholding parameter $\lambda=3$, and generate the $N\times p$ design
matrix with i.i.d. entries $\X_{i,j} \sim \mathcal{N}(0,\frac{1}{N})$. We fix the true
set of parameters $\thetastar$ to be a vector randomly drawn with fixed norm $\|
\thetastar \|_2 = 6 \sqrt{p}$, and then generate outcome $\yn$, $n=1,2,\ldots, N$, as
\begin{equation}
\yn = \xn^\top\thetastar + \epsilon_n.
\end{equation}
For the distribution of errors $\epsilon_n$, we use Huber's contaminated normal
distribution $\mathsf{CN}(0.05,10)$, i.e., $\epsilon_n\sim 0.95 z + 0.05
h_{10}$, i.i.d., where $z$ is standard normal and $h_x$ is a point mass at $x$.

Few alternative packages to \pkg{sgd} exist for high-dimensional robust
estimation. We compare to \pkg{hqreg} \citep{yi2015hqreg}, which fits
regularization paths for Huber loss regression with the elastic net penalty.
Note that \pkg{hqreg} is specialized to the Huber loss and cannot perform
estimation for the general setting of M-estimation problems considered here.

\begin{table}[!t]
  \begin{center}
  \begin{tabular}{llllll}
  \toprule
  $N$ & $p$ & \pkg{sgd} (\gls{AISGD}) & \pkg{sgd} (\gls{SGD}) & \pkg{hqreg} &
  units\\
  \midrule
  1,000 & 100 & 0.05 & 0.04 & 0.03 & (sec)\\
  10,000 & 500 & 0.55 & 0.46 & 0.40 & (sec)\\
  10,000 & 1,000 & 1.30 & 2.22 & 6.34 & (sec)\\
  50,000 & 10,000 & 3.12 & 3.86 & 15.57 & (min)\\
  100,000 & 50,000 & 8.13 & 15.20 & -- & (min)\\
  1,000,000 & 100,000 & 35.88 & 51.93 & -- & (min)\\
  10,000,000 & 100,000 & 8.64 & 9.55 & -- & (hr)\\
  100,000 & 1,000,000 & 18.80 & 26.43 & -- & (hr)\\
  \bottomrule
  \end{tabular}
  \end{center}
  \caption{High-dimensional M-estimation with the Huber loss. Timing (in units
  given by the last column) is displayed for 100 $\lambda$ values, averaged over
  10 runs. Omitted entries indicate failure of the algorithm; for \pkg{hqreg},
  it could not run due to memory limitations.}
  \label{table:huber}
\end{table}

Table \ref{table:huber} outlines results for a combination of pairs $(N,p)$,
ranging from small problems of $N=1,000$ observations to massive data settings
of $N=10$ million.
We apply the elastic net penalty with $\alpha=0.5$, which puts even weight on both
the lasso and ridge components, and then compute a regularization path for both
packages.
We also include an example of $N=100,000$ observations and $p=1,000,000$
covariates, where there exist far more covariates than data points; this occurs
often in applications, e.g., in text analysis, bioinformatics, and signal
processing \citep{lustig2008compressed,blei2012probabilistic}.

The \gls{SGD} algorithms begin to outperform \pkg{hqreg} on the order of tens of
thousands of observations, and significantly so for larger data settings.
Similar to the memory limitations of \pkg{glmnet}, \pkg{hqreg} requires access
to the full data set per iteration of its algorithm, which is infeasible when
the data cannot be held in memory. Thus we were unable to obtain proper timings
for data sets of size greater than $50,000$ observations and
$10,000$ covariates.

\begin{figure}[!tb]
  \begin{minipage}[b]{0.5\linewidth}
    \centering
    \includegraphics[width=\textwidth]{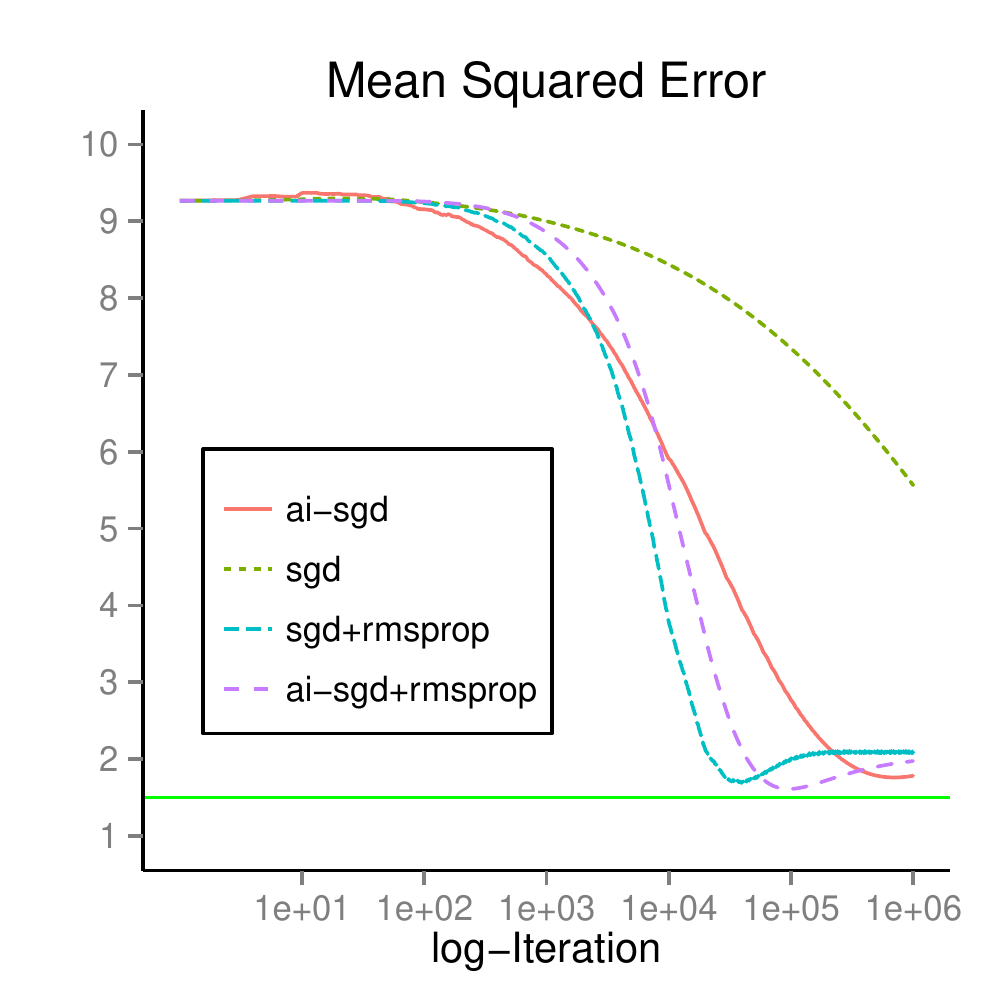}
    \phantomsubcaption \label{subfig:huber:1}
  \end{minipage}
  \hfill
  \begin{minipage}[b]{0.5\linewidth}
    \centering
    \includegraphics[width=\textwidth]{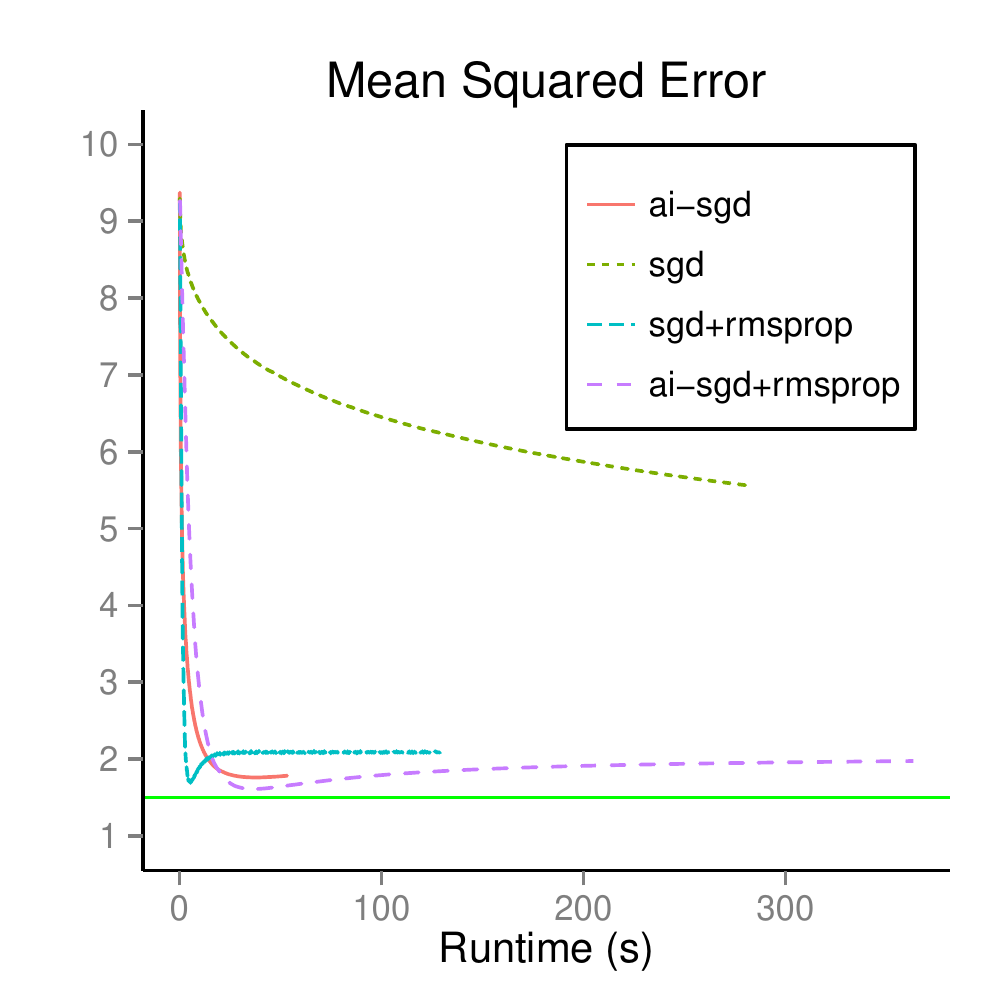}
    \phantomsubcaption \label{subfig:huber:2}
  \end{minipage}
  \caption{High dimensional M-estimation with the Huber loss, for $N=100,000$
  observations, $p=10,000$ covariates, and a fixed regularization parameter
  $\lambda$. The plots indicate the mean-squared error across iterations (left)
  and time (right) for \gls{SGD} algorithms. The horizontal line displays the
  mean-squared error for the exact $M$-estimator $\widehat\theta^{m}$.}
  \label{fig:huber}
\end{figure}

Figure \ref{fig:huber} displays the progress of the \gls{SGD} algorithms for the
setting of $N=100,000$ observations and $p=10,000$ covariates, for a fixed
regularization parameter $\lambda$. For demonstration, we run the algorithms
over 10 passes of the data and thus over a total of 1 million iterations.
\gls{AISGD} is seen to achieve a significantly faster convergence rate than
explicit \gls{SGD}. We also consider the use of adaptive schedules, here with
RMSProp, as it performs the fastest among other available learning rates
(see Section \ref{sec:interface:learning}). With RMSProp, the difference between the two
methods---\gls{SGD} and \gls{AISGD}---is noticeably smaller, and in fact
\gls{SGD} seems to converge slightly faster. We note however that the use of
\gls{SGD} algorithms with RMSProp breaks statistical efficiency, and indeed we
see this effect as the mean-squared error oscillates around a value higher than the
MSE of the exact M-estimator (green line). Therefore we advocate the use of
\gls{AISGD} with a one-dimensional learning rate, which still converges quite
quickly.

\section{Interface and implementation}
\label{sec:interface}
We now discuss the interface of \pkg{sgd} and various technical details
that are important for its use in practice.

\subsection{Interface}
The \pkg{sgd} package provides an intuitive and accessible set of methods
for performing estimation with large-scale data sets.
At the core of the package is the function
\begin{Code}
sgd(formula, data, model, model.control, sgd.control)
\end{Code}
The user provides a \code{formula} on the data frame
\code{data}---similar to function primitives, such as \code{lm}---and then specifies the
\code{model}. The model parameters are estimated using \gls{SGD} methods, which defaults to \gls{AISGD}. The optional arguments
\code{model.control} and \code{sgd.control} specify attributes one can tweak
about the model and the stochastic gradient method, respectively.
For example, given a data frame \code{dat} with response vector stored as the
column \code{y},
\begin{Code}
sgd(y ~ ., data=dat, model="lm")
\end{Code}
fits a linear model with the default specifications, e.g., \gls{AISGD} with a
one-dimensional learning rate.
Similarly,
\begin{Code}
sgd(y ~ ., data=dat, model="glm", model.control=list(family="binomial"))
\end{Code}
fits logistic regression with the default specifications.
Numerous examples are available in the package by running
\code{demo(package="sgd")}.

The \code{sgd} function also interfaces with data sets that are too large to fit
into memory or are streaming (more details in Section \ref{sec:interface:software}), and
 can be run with a custom loss function if desired.

The output of the \code{sgd} function is a \code{sgd} object, which is a light
wrapper on a \code{list} which collects quantities, such as the final parameter
estimates and convergence diagnostics. Custom generic methods are also
available for the \code{sgd} class, such as \code{print}, \code{predict}, and
\code{plot}.

\subsection{Stochastic gradient methods}
\label{sec:interface:stochastic}
While we describe the \gls{ESGD} and \gls{AISGD} algorithms in
Section \ref{sec:algorithms}, the following stochastic gradient methods are also
implemented in \pkg{sgd}:
\begin{itemize}
\item \gls{ISGD}: Proposed by \citet{toulis2014statistical} in the context
of generalized linear models, this
algorithm uses the implicit update \eqref{eq:implicit} and does not do any
iterate averaging.
\item \gls{ASGD}: Proposed by \citet{ruppert1988efficient} and
\citet{bather1989stochastic} independently, this algorithm uses the explicit
update \eqref{eq:explicit} followed by iterate averaging \eqref{eq:averaging}.
\item \gls{CM}: Proposed by \citet{polyak1964some}, this algorithm uses the update
\begin{align}
\label{eq:cm}
v_n &= \mu v_{n-1} + a_n \scoren{\theta_{n-1}}{n},\\
\theta_n &=  \theta_{n-1} + v_n,
\end{align}
where $\mu\in[0,1]$ is a fixed momentum coefficient. \gls{CM} accelerates
gradient descent with a velocity vector which accumulates directions of large
increase in the log-likelihood.
\item \gls{NAG}: Proposed by \citet{nesterov1983method}, this algorithm uses the update
\begin{align}
\label{eq:cm}
v_n &= \mu v_{n-1} + a_n \scoren{\theta_{n-1} + \mu v_{n-1}}{n},\\
\theta_n &=  \theta_{n-1} + v_n,
\end{align}
where $\mu\in[0,1]$ is a fixed momentum coefficient.  \gls{NAG} is similar to
\gls{CM} but accumulates velocity at a "look-ahead" point $\theta_{n-1}+\mu
v_{n-1}$. This makes a partial update closer to $\theta_n$, allowing \gls{NAG}
to change its velocity more quickly and responsively.
\end{itemize}
While all these methods are available, we recommend and apply \gls{AISGD} as the
default. It can be seen as an effective combination of the advantages from both
\gls{ISGD} and \gls{ASGD} \citep{toulis2015stability}. The momentum-based
methods \gls{CM} and \gls{NAG} enjoy faster convergence rates than the original
\gls{ESGD}, but offer no theoretical benefits against \gls{AISGD}. Without
averaging techniques they also are statistically inefficient, whereas iterate
averaging can be interpreted as an acceleration technique because
larger learning rates are used. The velocity
update in \gls{NAG} is also a proxy for the implicit update, as its benefit mostly
relies on making updates close to where the new estimate would lie.

\subsection{Learning rates}
\label{sec:interface:learning}
We describe the available learning rates in more detail because
they are critical for convergence of SGD methods, in practice.
It is well-known \citep{sakrison1965, amari1998natural,  toulis2014statistical}  that
\gls{ESGD} \eqref{eq:explicit} and \gls{ISGD} \eqref{eq:implicit}
have optimal statistical efficiency if the learning rate sequence $\gamma_n$ together, with the conditioning matrices $C_n$, approximate the inverse
Fisher information matrix $\Fisher{\thetastar} = - \Ex{\nabla^2 \ell(\thetastar; \xn, \yn)}$, i.e., $\gamma_n C_n \to \Fisher{\thetastar}^{-1}$, in the limit.
Therefore in first-order methods where $C_n=I$, the learning rate sequence acts as a scalar-valued approximation to
the optimal rescaling as it is used in Fisher scoring \citep{Fisher:1925vn}.
Based on this theory, the following learning rates are implemented in \pkg{sgd}:
\begin{itemize}
\item One-dimensional \citep{xu2011towards}: The learning rate is of the form
\begin{equation*}
\gamma_n =  \gamma_0(1 + a \gamma_0 n)^{-c},
\end{equation*}
where $\gamma_0, a, c\in\mathbb{R}$ are fixed constants. For \gls{SGD}
algorithms without iterate averaging and \gls{SGD} algorithms with iterate
averaging, \citet{xu2011towards} proved that setting $c=1$ and $c=2/3$,
respectively, leads to optimal statistical efficiency; a similar result
holds for \gls{AISGD} \citep{toulis2015stability}.
\item AdaGrad \citep{duchi2011adaptive}:
Rather than specify a one-dimensional learning rate $\gamma_n\in\mathbb{R}$, \citet{duchi2011adaptive} propose a diagonal conditioning matrix $C_n\in\mathbb{R}^{p\times p}$ given by
\begin{align*}
\mathcal{I}_n &= \mathcal{I}_{n-1} + \operatorname{diag}(\nabla\ell(\theta_{n-1}; \xn, \yn)\nabla\ell(\theta_{n-1}; \xn, \yn)^\top),\\
C_n &=  \eta(\mathcal{I}_n + \epsilon I)^{-1/2},
\end{align*}
where $\operatorname{diag}(\cdot)$ extracts the diagonal entries of its matrix argument, $\eta\in\mathbb{R}$ is a constant, $I$ is the identity matrix, and
$\epsilon$ is a fixed value, typically $10^{-6}$, to prevent division by zero. In the limit,
$I_n$ is an unbiased estimate of the diagonal entries of the Fisher
information, and the proposed diagonal matrix $C_n$, which accumulates such
curvature information, is proven to be optimal for minimization
of the regret bound.
\item RMSProp \citep{tieleman2012lecture}: A learning rate which is popular in the deep
learning literature \citep{srivastava2014dropout,ranganath2015deep,rezende2015variational},
\citet{tieleman2012lecture} propose the diagonal conditioning matrix $C_n\in\mathbb{R}^{p\times p}$ given by
\begin{align*}
\mathcal{I}_n &= \beta \mathcal{I}_{n-1} + (1 - \beta)\operatorname{diag}(\nabla\ell(\theta_{n-1}; \xn, \yn)\nabla\ell(\theta_{n-1}; \xn, \yn)^\top),\\
C_n &= \eta(\mathcal{I}_n + \epsilon I)^{-1/2},
\end{align*}
where $\beta\in[0,1]$ is the \emph{discount factor}, $\eta\in\mathbb{R}$ is a
constant, $I$ is the identity matrix, and $\epsilon$ is a fixed value to prevent division by zero, as in AdaGrad.
RMSProp uses a
decay in the estimate for the Fisher information by taking a weighted average,
and thus it gives more weight onto newer than older information.
RMSProp aims to offset one
problem AdaGrad often encounters in practice, where very large values occur for initial estimates of $I_n$ (e.g., due to poor initialization),
thus slowing down the AdaGrad procedure as it tries to accumulate enough curvature information to compensate for such an error \citep{schaul2014unit}.
RMSProp balances this by taking a weighted average of
previous and new information, and sees much empirical success.
One problem, however, is that RMSProp is no longer decaying sufficiently quickly \citep{robbins1951,duchi2011adaptive}, and thus it has no guarantees on
convergence. Moreover, assuming convergence, the limit of the learning rate
sequence is a constant, which makes the iterates jitter
around the true parameter value, ad infinitum.
\item Fisher: Following results on statistical efficiency and Fisher scoring,
we propose a learning rate using a diagonal conditioning matrix
$C_n\in\mathbb{R}^{p\times p}$ given by
\begin{align*}
\mathcal{I}_n &= (1-\gamma_n) \mathcal{I}_{n-1} + \gamma_n \mathrm{diag}(\nabla\ell(\theta_{n-1}; \xn, \yn)\nabla\ell(\theta_{n-1}; \xn, \yn)^\top),\\
C_n &= (\mathcal{I}_n + \epsilon I)^{-1},
\end{align*}
where $\gamma_n\propto1/n$, and $\epsilon$ is a small fixed value to prevent division by zero, as in AdaGrad.
As before, $I_n$ in the limit is an unbiased estimate of the diagonal Fisher
information, and $C_n$ is adaptive to curvature information.
\end{itemize}
One critical but often unnoticed issue with AdaGrad, RMSProp, and similar
adaptive schedules is that they are statistically inefficient: the
specification of the learning rates leads to biased estimation of the inverse
Fisher information matrix $\Fisher{\thetastar}^{-1}$ that, as mentioned earlier, is necessary for optimal statistical efficiency (an important exception is iterate averaging).
This leads to a suboptimal asymptotic variance for the \gls{SGD} procedure.
Thus we recommend and apply the last proposed learning rate (``Fisher'') by
default: it takes advantage of the curvature information such methods benefit
from, while still preserving as much statistical efficiency as possible in
diagonal conditioning matrices.


\subsection{Software integration}
\label{sec:interface:software}
For data sets that cannot be loaded into memory, we access subsets of the data
using \pkg{bigmemory} \citep{kane2013scalable}. This allows one to perform
stochastic gradient descent by passing over the data loaded into RAM, and then to
reload a new data set. This naturally applies to both large data sets, e.g., on
the order of dozens of gigabytes, and streaming settings, in which one has
access only to a subset of the (potentially infinite) data at a time.

In principle, with \pkg{bigmemory} the memory requirement for these stochastic
gradient methods is only a single data point and the current parameter estimate,
which is the minimum $\mathcal{O}(p)$ complexity for simply storing the estimate. In our
implementation we use these savings to try to load as much data into RAM as
possible. This speeds up convergence in practice, as it reduces
the amount of I/O overhead; this especially becomes a significant bottleneck when reading many objects from disk.

For fast implementations we use \pkg{Rcpp} \citep{eddelbuettel2011rcpp}, where
all algorithms are written in \proglang{C++} and only interface-level code is
written in \proglang{R}. Aside from the major computational gains, this also provides the opportunity to extend the library
to other programming languages. \pkg{RcppArmadillo} \citep{eddelbuettel2014rcpp}
is applied for access to pre-optimized linear algebra routines,
 and \pkg{BH} \citep{eddelbuettel2015bh} for access to the Boost libraries. We apply
template meta-programming and reusable classes in an object-oriented framework,
including concepts such as stochastic gradient methods, models, and learning rates.
Such concepts make it easy for other users to develop new algorithms and prototype them
in their own research or practices.

The plotting routines adopt many features from \pkg{ggplot2}
\citep{wickham2009ggplot2}, and are effectively templated \code{ggplot} objects.
Our software is also robust through unit
testing which follows the paradigm from \pkg{testthat} \citep{wickham2011testthat}.

\section{Discussion}
\label{sec:discussion}
%

As \gls{ESGD} has been used extensively in practice, particularly in the deep
learning community, many heuristics have been proposed to solve issues that
often occur. We describe several of these issues and their proposed solutions
in the literature, and compare to how our \pkg{sgd} package handles them.

\parhead{Overfitting.}
As \gls{SGD} algorithms simply minimize a loss function evaluted over the
training data, overfitting is a prevalent problem as it is for all
estimation methods. This is particularly an issue in complex
likelihood functions such as neural networks (see, e.g.,
\citet{giles2001overfitting,bakker2003task}).  Even with penalization terms
that try to offset the fit of the parameters, it is still difficult for
\gls{ESGD} to find the right set of hyperparameters for such regularization
without a computationally intensive search.

As a solution many practictioners adopt \emph{early stopping}, which simply
halts the optimization routine before it converges. However, there is little theory on
the estimates obtained from early stopping. Most practically, it is difficult to
know when to stop the algorithm and how to use it in combination with other
regularization techniques, such as penalization.

Fortunately, one of the advantages of \gls{AISGD} is that it requires less such
tweaking: the implicit update effectively performs a regularization as seen from
the Bayesian perspective, c.f., Section \ref{sec:introduction}. We've also seen in
practice that penalization terms do not affect the final estimates from
\gls{AISGD}, which makes it less reliant on heuristics, such as early stopping.

\parhead{Vanishing or exploding gradients.}
The numerical instability of \gls{ESGD} is a widespread issue in practice
\citep{bengio1994learning,hochreiter1998vanishing,hochreiter2001gradient, toulis2014statistical}. The stochastic gradients can easily be too large leading to
divergence, and when chained through compositions of functions can either vanish
to zero, or even explode to numerically infinite values;
for example, \citet{toulis2014statistical}
demonstrate the instability of \gls{ESGD} in a simple bivariate Poisson model,
where slight misspecification of learning rate parameters lead to divergence.

\citet{pascanu2012difficulty} propose \emph{gradient
clipping}, which simply thresholds the stochastic gradient if it is outside a
bounded interval. Unfortunately, while it can work in practice, it is a
heuristic that breaks the key assumptions for convergence rate guarantees on
\gls{SGD} algorithms. Similarly, there is no principled way to set the bounds.
For \gls{AISGD} algorithms applied to the settings we consider in
Section \ref{sec:algorithms}, such an issue never arises. Theoretical results
establish stability regardless of the specification of the learning rate \citep[Section 3]{toulis2015implicit}, and
perform well in practice, as seen in Section \ref{sec:experiments}.


\section{Concluding remarks}
The \pkg{sgd} package is the most extensive implementation in \proglang{R} of
stochastic gradient methods for estimation with massive and/or streaming data sets.
Thus, \pkg{sgd} broadens the capabilities of
\proglang{R} for estimation with modern large data sets---on the orders of
hundreds of millions of observations and hundreds of thousands of
covariates---while retaining desirable statistical properties.
The software is based on solid theory of stochastic approximations, which help
guide the optimal selection of parameters, e.g., learning rates, in the
underlying optimization routines.
In this paper, we show how \pkg{sgd} can be applied for estimation of
\acrlongpl{GLM} and \glslink{M}{M-estimation}, which comprise a
sizeable portion of estimation problems encountered in statistical practice.

There are many software extensions that are currently in development.  We
are working to interface with other high-performance computing packages,
namely \pkg{sqldf} for faster I/O applications with streaming data, \pkg{doParallel}
\citep{revolution2014doparallel} and \pkg{Rmpi} \citep{yu2002rmpi} for parallel
updates across environments, and \pkg{gputools} \citep{buckner2011gputools} for
efficient computing with GPUs. The algorithms described here directly appeal to
asynchronous implementations, following \code{Hogwild!} \citep{nui2011hogwild},
which allows for lock-free allocation of CPU cores. Sparse data structures would
allow for fast structured matrix and vector products, which occur, for example,
when looping over the covariates of a data point, and would significantly speed
up computation on sparse data sets.

Finally, there has been little attention on, and in fact a pressing need for,
model selection and hypothesis testing in \gls{SGD} procedures. We are pursuing this
in light of the new statistical challenges presented to us while developing the
\pkg{sgd} package.


\bibliography{jss}

\end{document}